\documentclass[aps,prl,superscriptaddress,floatfix, twocolumn]{revtex4}

\usepackage{graphicx,graphics}
\usepackage{dcolumn}
\usepackage{amsmath,amssymb,amsfonts}
\usepackage{latexsym,verbatim}
\usepackage{bm}
\usepackage{color}
\usepackage{ulem}

\def\be{\begin{equation}}
\def\ee{\end{equation}}
\def\ber{\begin{eqnarray}}
\def\eer{\end{eqnarray}}

\begin{document}
\title{Acoustic plasmons and ``soundarons" in graphene on a metal gate}
\author{A. Principi}
\affiliation{NEST, Istituto Nanoscienze-CNR and Scuola Normale Superiore, I-56126 Pisa, Italy}
\author{Reza Asgari}
\email{asgari@ipm.ir}
\affiliation{School of Physics, Institute for Research in Fundamental Sciences (IPM), Tehran 19395-5531, Iran}
\author{Marco Polini}
\email{m.polini@sns.it}
\affiliation{NEST, Istituto Nanoscienze-CNR and Scuola Normale Superiore, I-56126 Pisa, Italy}
\begin{abstract}
We demonstrate that single-layer graphene in the presence of a metal gate displays a gapless collective (plasmon) mode that has a linear dispersion at long wavelengths. We calculate exactly the acoustic-plasmon group velocity at the level of the random phase approximation and carry out microscopic calculations of the one-body spectral function of such system. Despite screening exerted by the metal, we find that graphene's quasiparticle spectrum displays a very rich structure characterized by composite hole-acoustic plasmon satellite bands (that we term for brevity ``soundarons"), which can be observed by {\it e.g.} angle-resolved photoemission spectroscopy.
\end{abstract}

\maketitle

{\it Introduction. ---} Graphene, a monolayer of carbon atoms packed in a two-dimensional (2D) honeycomb lattice, is a gapless semiconductor whose recent isolation has generated an enormous deal of interest in various fields of condensed matter physics~\cite{graphenereviews}. Electrons in graphene move as if they were massless Dirac fermions (MDFs) with a velocity $v$ which is roughly three-hundred times smaller than the velocity of light in vacuum. Apart from a few notable exceptions~\cite{earlymanybodyeffects}, until 2009 many electronic and optical properties of graphene could be explained within a single-particle picture in which electron-electron (e-e) interactions are completely neglected. The discovery of the fractional quantum Hall effect in graphene~\cite{fqhe} represents an important  hallmark in this context. By now there is a large body of experimental work~\cite{eeinteractionsgraphene,bostwick_science_2010,kotov_arXiv_2010} showing the relevance of e-e interactions in a number of key properties of graphene samples of sufficiently high quality. 

Recently, it has been demonstrated that the MDF energy spectrum of graphene is significantly altered by many-particle effects: robust ``{\it plasmaron}" satellite bands have been first predicted~\cite{polini_prb_2008,hwang_prb_2008} and then observed~\cite{bostwick_science_2010} in quasi-freestanding graphene grown on hydrogen-terminated SiC. Plasmarons~\cite{earlywork} are composite quasiparticles that emerge in an electron liquid when the interactions between electrons and plasmons -- the collective density oscillations of the electron liquid~\cite{Giuliani_and_Vignale} -- are particularly strong. A deep understanding of the coupling between charge carriers and plasmons in graphene is highly desirable because of the large potential of this material~\cite{grapheneplasmonics} in the context of plasmonics~\cite{plasmonics}.  

This Letter is motivated by the large amount of work that has been carried out by many groups around the world~\cite{grapheneonmetals} to grow high quality graphene sheets on transition-metal templates such as Ru, Ir, Ni, Pt, Cu, and Au. Depending on the metal, there can be weak or strong hybridization between graphene $\pi$ and metal $d$ bands. In the former case, electrons close to the Fermi energy in the graphene sheet are still described by a MDF Hamiltonian: the main qualitative role of the metal is to screen e-e interactions between MDFs, weakening them. 
In this Letter we present a careful analytical and numerical analysis of the impact of a metal gate on the collective behavior of the electron gas in a nearby graphene sheet: we show that, contrary to expectation, screened e-e interactions in graphene on a metal gate (G/M) can lead to intriguing  many-body effects and rich spectral features.

{\it Model Hamiltonian and acoustic plasmons. ---} We model the electron gas in G/M by the following continuum Hamiltonian (hereby written for a single channel): 
\begin{equation}\label{eq:MDFhamiltonian}
{\hat {\cal H}} =  \hbar v \sum_{{\bm k}, \alpha, \beta} {\hat \psi}^\dagger_{{\bm k}, \alpha} 
( {\bm \sigma}_{\alpha\beta} \cdot {\bm k} ) {\hat \psi}_{{\bm k}, \beta} + \frac{1}{2 S}\sum_{{\bm q} \neq {\bm 0}} V_d(q){\hat \rho}_{\bm q} {\hat \rho}_{-{\bm q}}~.
\end{equation}
Here $v$ is the MDF velocity, $S$ is the sample area, and
$
{\hat \rho}_{\bm q} = \sum_{{\bm k}, \alpha} {\hat \psi}^\dagger_{{\bm k} - {\bm q}, \alpha}{\hat \psi}_{{\bm k}, \alpha}
$
is the density operator. Greek letters are honeycomb-sublattice-pseudospin labels and ${\bm \sigma} = (\sigma^x,\sigma^y)$ is a 2D vector of Pauli matrices. 
Since we are interested only in the role of screening played by the metal gate, in writing Eq.~(\ref{eq:MDFhamiltonian}) we have neglected hybridization i) between the graphene sheet and the metal and ii) between the surface plasmons of the metal~\cite{surfaceplasmons} and the low-energy ``sheet" plasmon of the graphene electron gas. The metal is thereby modeled as a {\it grounded conductor} parallel to the graphene sheet. The function $V_d(q)$ represents the Fourier components of the e-e interaction, altered by the presence of the grounded gate~\cite{raoux_prb_2010}:
$
V_d(q) = 2\pi e^2 [1 - \exp(-2 q d)]/q,
$
$d$ being the graphene-metal gate distance. Due to screening exerted by the metal, $V_d(q)$ is regular at $q =0$ for any finite $d$, {\it i.e.} $V_d(q \to 0) = 4\pi e^2 d$. In the limit $d \to \infty$ one recovers the usual 2D Fourier transform of the long-range Coulomb potential, {\it i.e.} $V_\infty(q) = 2\pi e^2/q$.  

Whether or not the model represented by Eq.~(\ref{eq:MDFhamiltonian}) is adequate to describe the electron gas in epitaxial graphene grown on metal substrates~\cite{grapheneonmetals} (the metal being one of those that hybridizes weakly with graphene such as Pt or Cu) can only be determined {\it a posteriori} by checking the predictions of this Letter against experiments. Our results are also relevant, though, for mechanically-exfoliated flakes in the proximity of a metal gate.

The collective density excitations of the Hamiltonian ${\hat {\cal H}}$ are completely determined by the density-density linear-response function,
$
\chi_{\rho\rho}(q,\omega) = \langle \langle {\hat \rho}_{\bm q}; {\hat \rho}_{-{\bm q}}\rangle\rangle_\omega/S,
$
with $\langle\langle {\hat A},{\hat B}\rangle\rangle_\omega$ the usual Kubo product~\cite{Giuliani_and_Vignale}.  Within the random phase approximation (RPA)~\cite{Giuliani_and_Vignale,commentRPA} $\chi_{\rho\rho}(q,\omega)$ is given by
\begin{equation}\label{eq:chiRPA}
\chi_{\rho\rho}(q,\omega) = \frac{\chi_0(q,\omega)}{1- V_d(q)\chi_0(q,\omega)} \equiv \frac{\chi_0(q,\omega)}{\varepsilon(q,\omega)}~,
\end{equation}
where $\chi_0(q,\omega)$ is the well-known~\cite{wunsch_njp_2006} response function of a 2D noninteracting gas of MDFs at arbitrary doping $n$. For future purposes, we also introduce the coupling constant $\alpha_{\rm ee} = e^2/(\hbar v)$, the Fermi wave number $k_{\rm F} = \sqrt{4 \pi |n|/(g_{\rm s} g_{\rm v})}$, $g_{\rm s} =2$ ($g_{\rm v} =2$) being a spin (valley) degeneracy factor, and the Fermi energy $\varepsilon_{\rm F} = v k_{\rm F}$. The collective modes of the Hamiltonian ${\hat {\cal H}}$ are particle-hole symmetric since $\chi_0(q,\omega)$ is so.

\begin{figure}[t]
\centering
\includegraphics[width=1.00\linewidth]{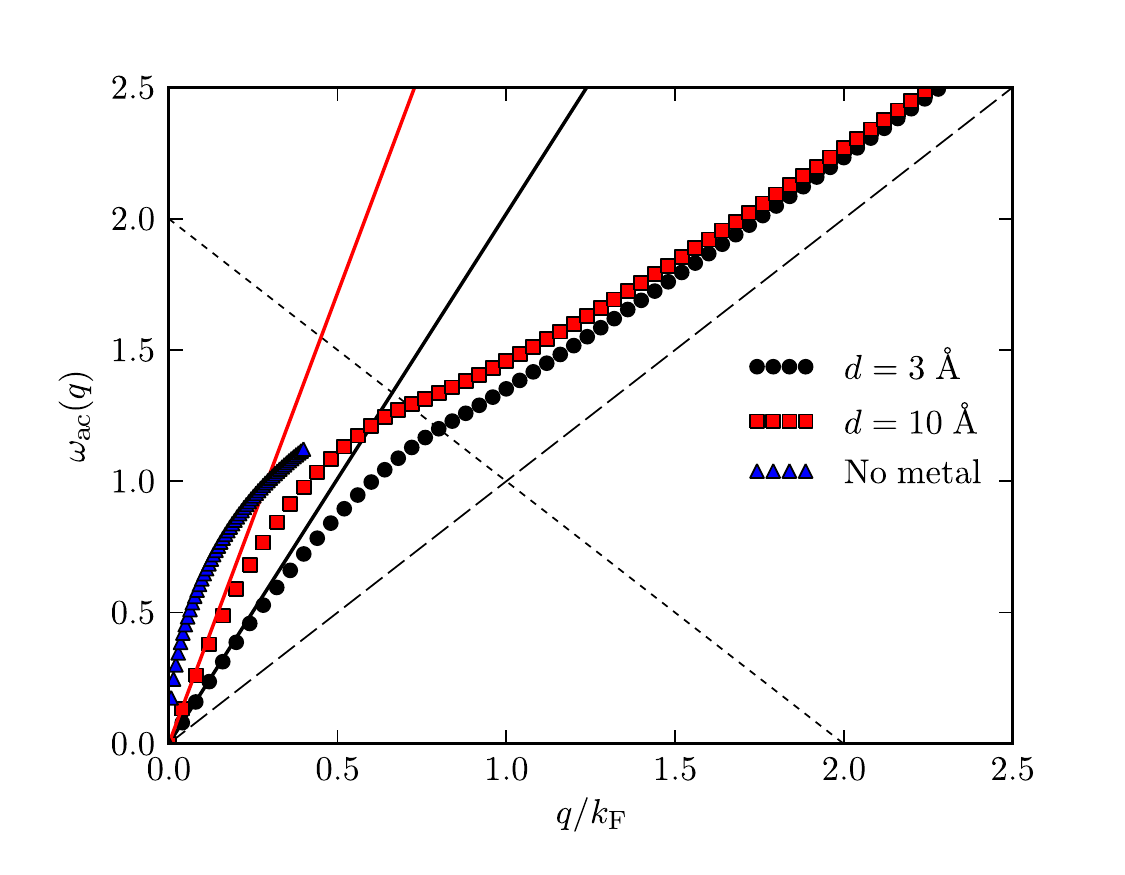}
\caption{(Color online) Acoustic plasmon dispersion (in units of $\varepsilon_{\rm F}$) in G/M. Filled circles and squares refer to different values of the graphene-metal gate distance $d$. Filled triangles refer to the standard plasmon ($\propto \sqrt{q}$ for $q \to 0$) in the absence of a metal gate, here plotted only up to $q = 0.4~k_{\rm F}$ for clarity. These results are for $|n| = 5.0 \times 10^{13}~{\rm cm}^{-2}$ and $\alpha_{\rm ee} = 2.2$. The long-dashed line ($\omega = vq$) represents the upper bound of the intra-band electron-hole continuum. The intersection between the acoustic plasmon and the thin short-dashed line ($\omega = 2 \varepsilon_{\rm F} - vq$, which is the lower bound of the inter-band electron-hole continuum) gives the critical wave number $q_{\rm c}$ at which Landau damping starts. Note that, for every fixed $q$, the energy of the acoustic plasmon decreases with decreasing $d$, since the screening power of the metal becomes larger at smaller values of $d$. At large $q$ the role of the metal gate is negligible and all plasmon curves collapse onto each other.
\label{fig:one}}
\end{figure}

The dispersion $\omega_{\rm ac} = \omega_{\rm ac}(q)$ of the collective (plasmon) mode can be found by solving the complex equation
$
\varepsilon(q,\omega) = 0
$
in the regions of the $q,\omega$ plane where $\Im m~[\chi_0(q,\omega)] =0$. Since $V_d(q)$ is regular at $q=0$, we expect a gapless {\it acoustic} plasmon, $\omega_{\rm ac}(q \to 0) =c_{\rm s} q$, rather than the usual (``unscreened") plasmon~\cite{wunsch_njp_2006,polini_prb_2008} $\propto \sqrt{q}$ for $q \to 0$. The latter behavior is formally recovered only at $d = \infty$. We now proceed to derive an exact expression for the RPA group velocity $c_{\rm s}$. Following Santoro and Giuliani~\cite{santoro_prb_1988}, we first introduce the power expansion
$
\omega_{\rm ac}(q) = c_{\rm s} q + c_2 q^2 + c_3 q^3 + \dots,
$
for the acoustic-plasmon dispersion, and then define the function
$
F(q) \equiv \varepsilon(q,c_{\rm s} q + c_2 q^2 + c_3 q^3 + \dots).
$
In the long-wavelength limit $F(q)$ admits the following Laurent-Taylor expansion
$
F(q \to 0) = f_0 + f_1~q + f_2~q^2 + \dots~,
$
where the coefficients $f_i$ can be derived from the analytical expression~\cite{wunsch_njp_2006} for $\chi_0(q,\omega)$. For $\varepsilon(q,\omega) =0$ to be valid we have to require that the coefficients $f_i$ vanish identically. The coefficient $f_0$ depends {\it only} on $c_{\rm s}$ and by equating its expression to zero we arrive after some tedious but straightforward algebra at the desired expression for the acoustic-plasmon group velocity $c_{\rm s}$:
\begin{equation}\label{eq:sound}
c_{\rm s} = v \sqrt{\frac{\Lambda(\alpha_{\rm ee}dk_{\rm F})}{\Lambda(\alpha_{\rm ee}dk_{\rm F})-1}}~,
\end{equation}
with $\Lambda(x) =[1 +1/(2 g_{\rm s} g_{\rm v} x)]^2 >1$. Note that the ``sound" velocity 
$c_{\rm s}$ is larger than $v$ for any value of the dimensionless product $\alpha_{\rm ee} d k_{\rm F}$. This implies that, within RPA, an undamped acoustic plasmon is {\it always} present at small $q$ in the system modeled by Eq.~(\ref{eq:MDFhamiltonian}). Eq.~(\ref{eq:sound}) is the most important analytical result of this work and provides a simple expression for the sound velocity that can be compared with on-going and future experiments on plasmons in G/M. In Fig.~\ref{fig:one} we present a comparison between the dispersion $\omega_{\rm ac}(q)$ of the acoustic plasmon as found from the numerical solution of $\varepsilon(q,\omega)=0$ and the long-wavelength analytical result $\omega_{\rm ac}(q \to 0) = c_{\rm s}q$, with $c_{\rm s}$ given by Eq.~(\ref{eq:sound}) (thin solid lines in Fig.~\ref{fig:one}): the agreement is clearly excellent.

\begin{figure}
\centering
\includegraphics[width=1.00\linewidth]{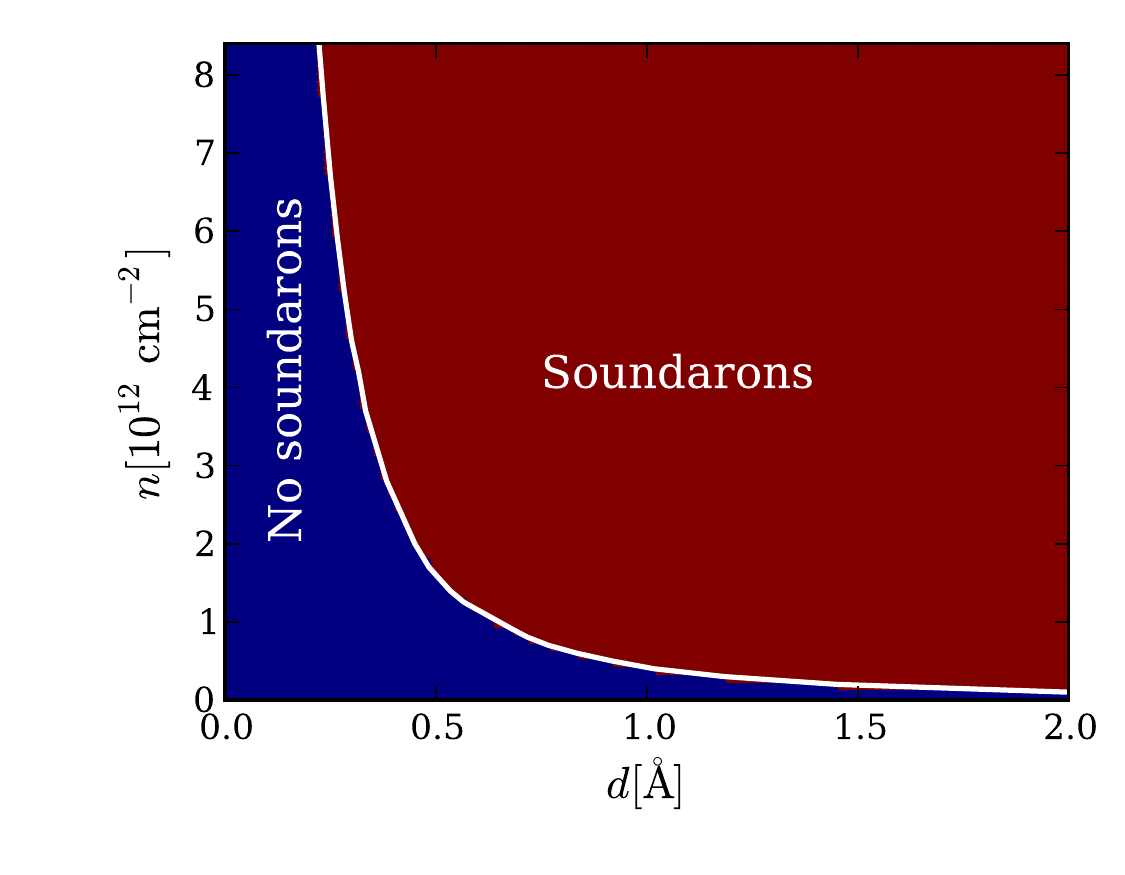}
\caption{(Color online) The  color red (blue) labels the region of parameter space in which resonant interactions between carriers and acoustic plasmons exist (do not exist) giving rise to ``soundaron" satellite bands in the spectral function ${\cal A}({\bm k}, \omega)$. This phase diagram is for a coupling constant $\alpha_{\rm ee} = 2.2$.
\label{fig:two}}
\end{figure}

The existence of acoustic plasmons has a large impact on the wave-vector and energy dependence of the one-body spectral function ${\cal A}({\bm k}, \omega)$~\cite{Giuliani_and_Vignale} of G/M. As explained in Refs.~\onlinecite{bostwick_science_2010} and~\onlinecite{polini_prb_2008}, {\it resonant} interactions between charge carriers and plasmons give rise to strong plasmaron satellite bands, which are clearly visible in ${\cal A}({\bm k}, \omega)$. In the system modeled by Eq.~(\ref{eq:MDFhamiltonian}), resonances occur when the 
bare quasiparticle velocity $v$ equals the acoustic-plasmon group velocity, {\it i.e.} when $v - \left.d\omega_{\rm ac}(q)/dq\right|_{q = q^\star} =0$, at some specific wave number $q^\star$. If $q^\star$ is smaller than the critical wave number $q_{\rm c}$ at which inter-band Landau damping starts (see Fig.~\ref{fig:one}), then a composite carrier-acoustic plasmon excitation, which we term for brevity ``soundaron", emerges. Despite its simplicity, the model in Eq.~(\ref{eq:MDFhamiltonian}) admits a non-trivial ``phase diagram" for the existence of soundarons, which is shown in Fig.~\ref{fig:two}. We clearly see that soundarons exist in a very broad region of parameter space. When resonances occur, strong satellite bands appear in the one-particle Green's function ${\cal A}({\bm k}, \omega)$.

{\it Quasiparticle decay rate and the one-body spectral function. ---} The spectral function ${\cal A}=  {\cal A}_+ + {\cal A}_-$ can be written in terms of the quasiparticle self-energy $\Sigma_\lambda({\bm k}, \omega)$~\cite{Giuliani_and_Vignale}:
\begin{equation}
{\cal A}_\lambda = \frac{\pi^{-1}|\Im m~[\Sigma_\lambda]|}{(\omega - \xi_{{\bm k}, \lambda} - \Re e~[\Sigma_\lambda])^2+ (\Im m~[\Sigma_\lambda])^2}~,
\end{equation}
where $\xi_\lambda({\bm k})= \lambda v k - \varepsilon_{\rm F}$ are Dirac band energies. In the so-called ``$G_0W$-RPA" approximation~\cite{bostwick_science_2010,Giuliani_and_Vignale,polini_prb_2008,hwang_prb_2008} the imaginary part of $\Sigma_\lambda({\bm k}, \omega)$ is given by the following expression:
\begin{eqnarray}\label{eq:res_computation}
\Im m~[\Sigma_\lambda({\bm k},\omega)] &=& \sum_{\lambda'} 
\int \frac{d^2 {\bm q}}{(2\pi)^2}~V_d(q)~\Im m[\varepsilon^{-1}({\bm q},\Omega)] \nonumber\\
&\times& {\cal F}_{\lambda,\lambda'}({\bm k}, {\bm k}^+)
\left[\Theta(\Omega)-\Theta(-\xi_{\lambda'}({\bm k}^+)) \right] \nonumber\\
\end{eqnarray}
where ${\bm k}^+ \equiv {\bm k}+{\bm q}$, $\Omega \equiv \omega-\xi_{\lambda'}({\bm k}^+)$, and 
$ {\cal F}_{\lambda,\lambda'} \equiv [1+ \lambda\lambda'\cos{(\theta_{{\bm k},{\bm k}^+})}]/2$. In Eq.~(\ref{eq:res_computation}) 
$\Theta(x)$ is the Heaviside step function. $ \Im m~[\Sigma_\lambda]$ measures the band-quasiparticle decay rate.
The two factors in the second line of Eq.~(\ref{eq:res_computation}) express the 
influence of chirality and Fermi statistics on the decay process, respectively. The ``quality" of the $G_0W$-RPA approximation for doped graphene sheets has been carefully tested in Ref.~\onlinecite{bostwick_science_2010} against angle-resolved photoemission spectroscopy (ARPES) data for graphene on SiC. Theory compares very well with the latter in a wide range of parameters~\cite{bostwick_science_2010}, thus demonstrating that $G_0W$-RPA is a very good starting point for doped graphene.

\begin{figure}[t]
\begin{center}
\tabcolsep=0cm
\begin{tabular}{cc}
\includegraphics[width=0.50\linewidth]{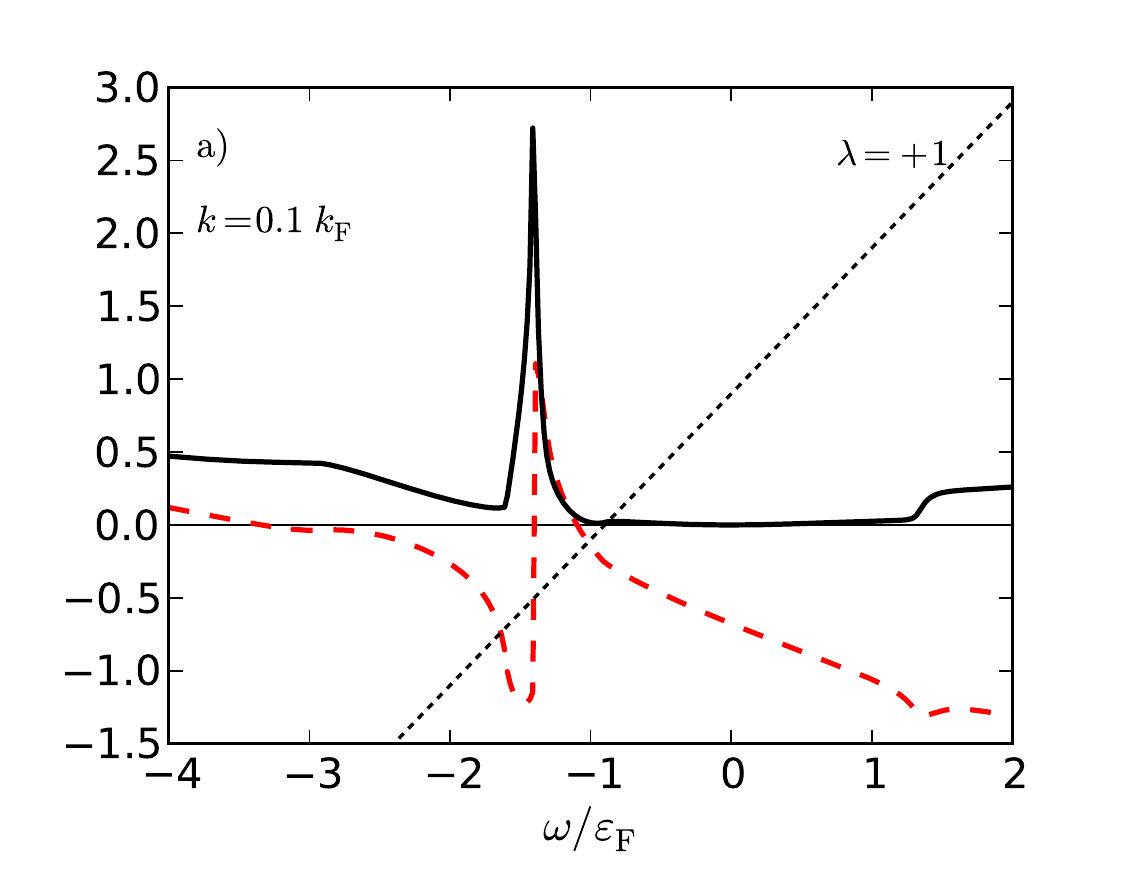}&
\includegraphics[width=0.50\linewidth]{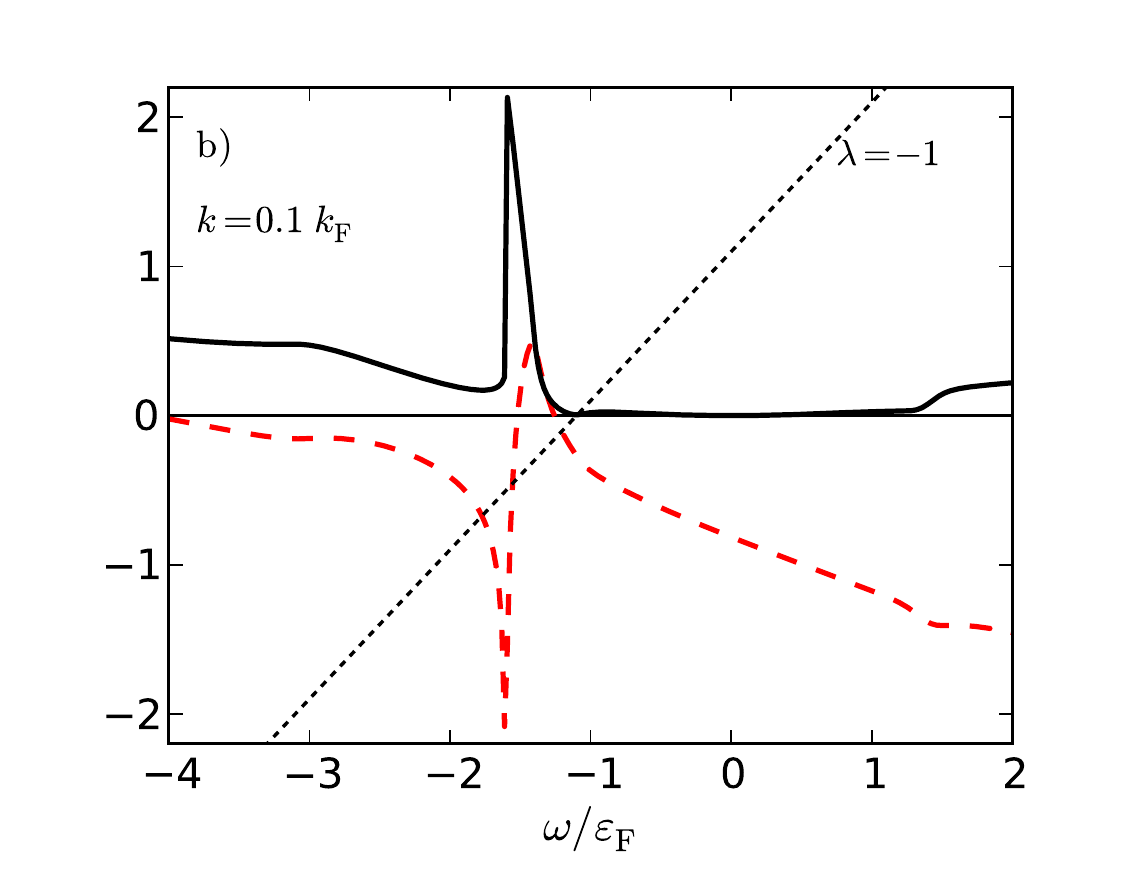}
\end{tabular}
\caption{(Color online) Quasiparticle decay rates and energies for $n$-doped G/M. Panel a) 
$|\Im m [\Sigma_+({\bm k}, \omega)]|$ (thick solid line) and $\Re e~[\Sigma_+({\bm k}, \omega)]$ (dashed line) as functions of energy $\omega$ (in units of and measured from the Fermi energy $\varepsilon_{\rm F}$) for $k=0.1~k_{\rm F}$. 
The parameters used in these plots are $n = 5.0 \times 10^{13}~{\rm cm}^{-2}$, $d = 5~{\rm \AA}$, and $\alpha_{\rm ee} = 2.2$. The intersections between $\omega - \xi_{{\bm k},+}$ (dotted line) and $\Re e~[\Sigma_+({\bm k}, \omega)]$ (dashed line) represent the solutions of the Dyson equation, $\omega - \xi_{{\bm k},+} - \Re e~[\Sigma_+({\bm k}, \omega)] = 0$. Panel b) Same as in panel a) but for $|\Im m [\Sigma_-({\bm k}, \omega)]|$ and $\Re e~[\Sigma_-({\bm k}, \omega)]$.\label{fig:three}}
\end{center}
\end{figure}

\begin{figure}
\centering
\includegraphics[width=1.00\linewidth]{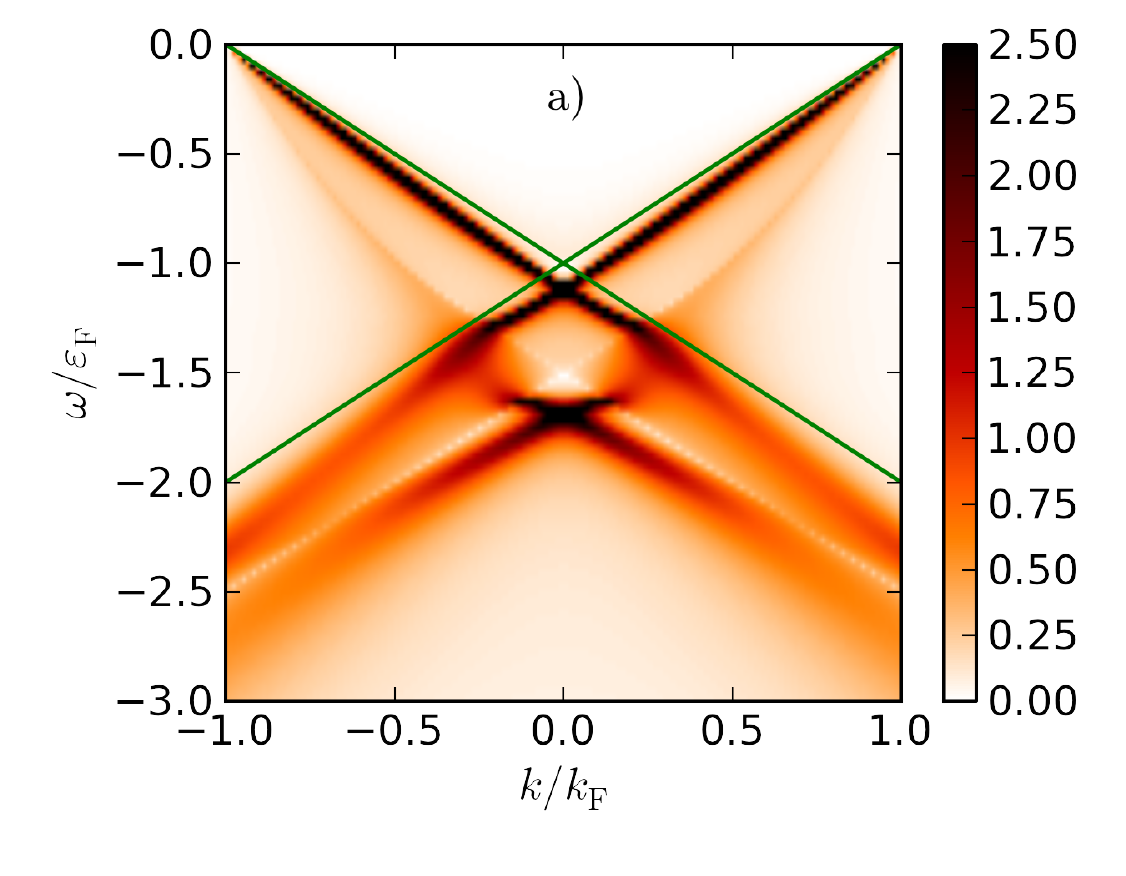}\\
\includegraphics[width=1.00\linewidth]{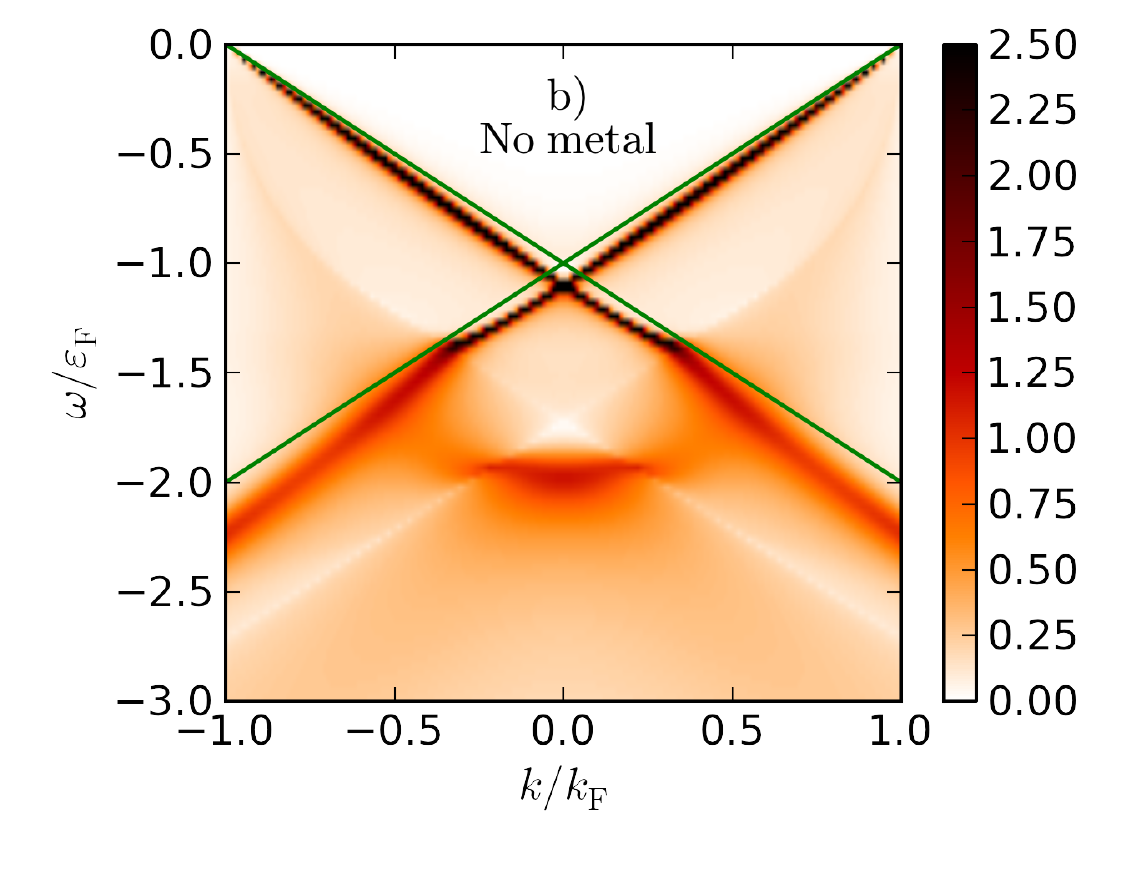}
\caption{(Color online) Panel a) Spectral function ${\cal A}({\bm k},\omega)$ of $n$-doped G/M. Energy $\omega$ is in units of and measured from $\varepsilon_{\rm F}$. This result is for $n = 5.0 \times 10^{13}~{\rm cm}^{-2}$, $d =5~{\rm \AA}$, and $\alpha_{\rm ee} = 2.2$. The solid lines represent the noninteracting bands. Panel b) Same as in panel a) but in the absence of the metal gate. The data in this panel have been obtained with the same theory and codes used in Refs.~\onlinecite{bostwick_science_2010} and~\onlinecite{polini_prb_2008}. The only difference is that the results in these earlier articles are for smaller values of the coupling constant $\alpha_{\rm ee}$ ($\leq 0.75$).
\label{fig:four}}
\end{figure}

Representative numerical results for $\Im m~[\Sigma_\lambda({\bm k},\omega)]$ and $\Re e~[\Sigma_\lambda({\bm k},\omega)]$ in the presence of a grounded metal gate are collected in Fig.~\ref{fig:three}. The main features we want to highlight in these plots are the sharp resonances in the quasiparticle decay rates stemming from resonant carrier-acoustic plasmon interactions, as discussed above. Since $\Re e~[\Sigma_\lambda({\bm k},\omega)]$ is related to $\Im m~[\Sigma_\lambda({\bm k},\omega)]$ by a Kramers-Kronig transform, sharp resonances in $\Im m~[\Sigma_\lambda({\bm k},\omega)]$ imply rapid changes in the real part of the quasiparticle self-energy and thus multiple solutions of the Dyson equation, $\omega - \xi_{{\bm k}, \lambda} - \Re e~[\Sigma_\lambda({\bm k},\omega)] =0$. Those solutions to which a small $\Im m~[\Sigma_\lambda({\bm k},\omega)]$ is associated give rise to sharp soundaron satellites in ${\cal A}({\bm k}, \omega)$. In Fig.~\ref{fig:four}a) we present the dependence of the spectral function ${\cal A}({\bm k}, \omega)$ on momentum and energy (below the Fermi energy) for $n$-doped G/M. We clearly see soundaron satellite bands, which are especially strong close to the Dirac point (${\bm k} = {\bm 0}$). The dependence of ${\cal A}({\bm k}, \omega)$ on $\alpha_{\rm ee}$ and $d$ is illustrated in the supplementary material. For the sake of comparison, in Fig.~\ref{fig:four}b) we have reported ${\cal A}({\bm k}, \omega)$ for a graphene sheet with the same doping and at the same coupling constant ($\alpha_{\rm ee} =2.2$) but in the absence of a metal gate. Contrary to expectation, the plasmaron satellite bands in Fig.~\ref{fig:four}b) are less pronounced than the soundaron satellite bands in G/M. This is in agreement with Ref.~\onlinecite{bostwick_science_2010}, in which it was clearly shown that plasmaron satellite bands become less intense and sharp at strong coupling ({\it i.e.} at large values of $\alpha_{\rm ee}$).

In summary, we have shown that screened e-e interactions in a graphene sheet deposited on a metal gate are responsible for intriguing many-body effects. Despite the strong screening exerted by the metal, the electron gas in the graphene sheet displays self-sustained, long-lived oscillations which have a linear dispersion at long wavelengths. These modes yield composite carrier-acoustic plasmon quasiparticle satellite bands in the one-particle Green's function.  Acoustic plasmons can be observed {\it e.g.} by inelastic light scattering or electron energy loss spectroscopy~\cite{liu_prb_2008,surfaceplasmons}. Soundarons can be detected in tunneling density-of-states~\cite{murphy_prb_1995} and angle-resolved photoemission spectroscopy~\cite{bostwick_science_2010,damascelli_rmp_2003} experiments. 

While this manuscript was being finalized, we learned of an {\it ab-initio} study~\cite{yan_prl_2011} that emphasizes the large impact of screening on the (high-energy) $\pi$ plasmon of graphene. Our work is complementary to this study since we focus entirely on the low-energy ``sheet" plasmon and, most importantly, on the spectral function. 

We thank D. Pacil\'e, M. Papagno, A. Politano, and E. Rotenberg for fruitful discussions and correspondence. M.P. acknowledges the kind hospitality of the IPM (Tehran, Iran) during the final stages of the preparation of this work.

\begin{appendix}
\section{SUPPLEMENTARY INFORMATION}

In this Section we report a series of numerical results to illustrate the dependence of the one-particle spectral function ${\cal A}({\bm k}, \omega)$ on the graphene-metal gate distance $d$ and fine-structure coupling constant $\alpha_{\rm ee}$. For the sake of completeness, we also report data for the spectral function of an isolated graphene sheet for different values of $\alpha_{\rm ee}$.

\begin{figure}[t]
\centering
\includegraphics[width=1.00\linewidth]{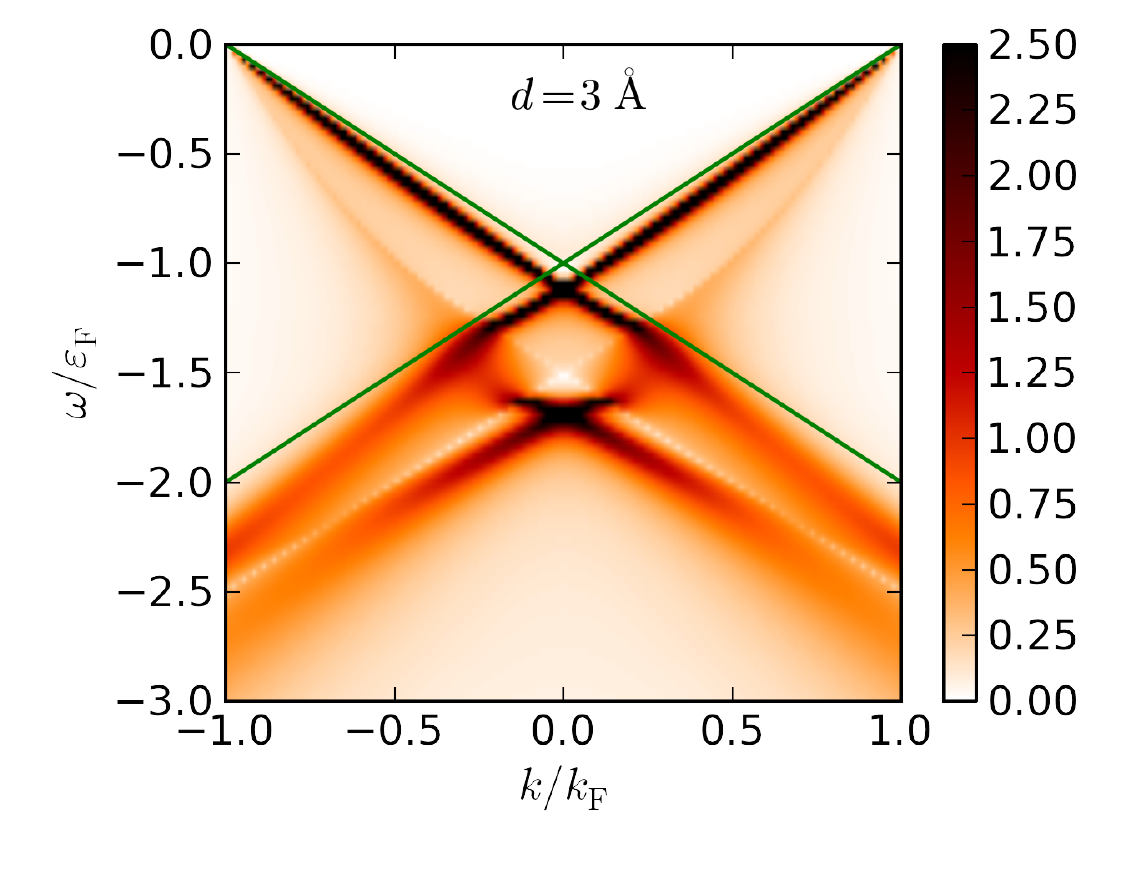}
\caption{(Color online) Same as in Fig.~\ref{fig:four}a) of the main text but for $d =3~{\rm \AA}$.
\label{fig:sm_1}}
\end{figure}

In Fig.~\ref{fig:sm_1} we show the $G_0W$-RPA spectral function ${\cal A}({\bm k}, \omega)$ of the electron gas in G/M for $n = 5.0 \times 10^{13}~{\rm cm}^{-2}$, $d =3~{\rm \AA}$, and for $\alpha_{\rm ee} = 2.2$. This result should be compared with that reported in Fig.~\ref{fig:four}a) of the main text, which was obtained by using a larger value of the graphene-metal substrate distance 
({\it i.e.} $d = 5~{\rm \AA}$). We clearly see that the soundaron satellite bands are robust against changes in $d$. This is in agreement with the phase diagram presented in Fig.~\ref{fig:two} of the main text: for large values of the carrier concentration ($n \gtrsim 10^{13}~{\rm cm}^{-2}$) the critical distance $d$ at which soundarons disappear is truly minute.

\begin{figure}[h!]
\centering
\includegraphics[width=1.00\linewidth]{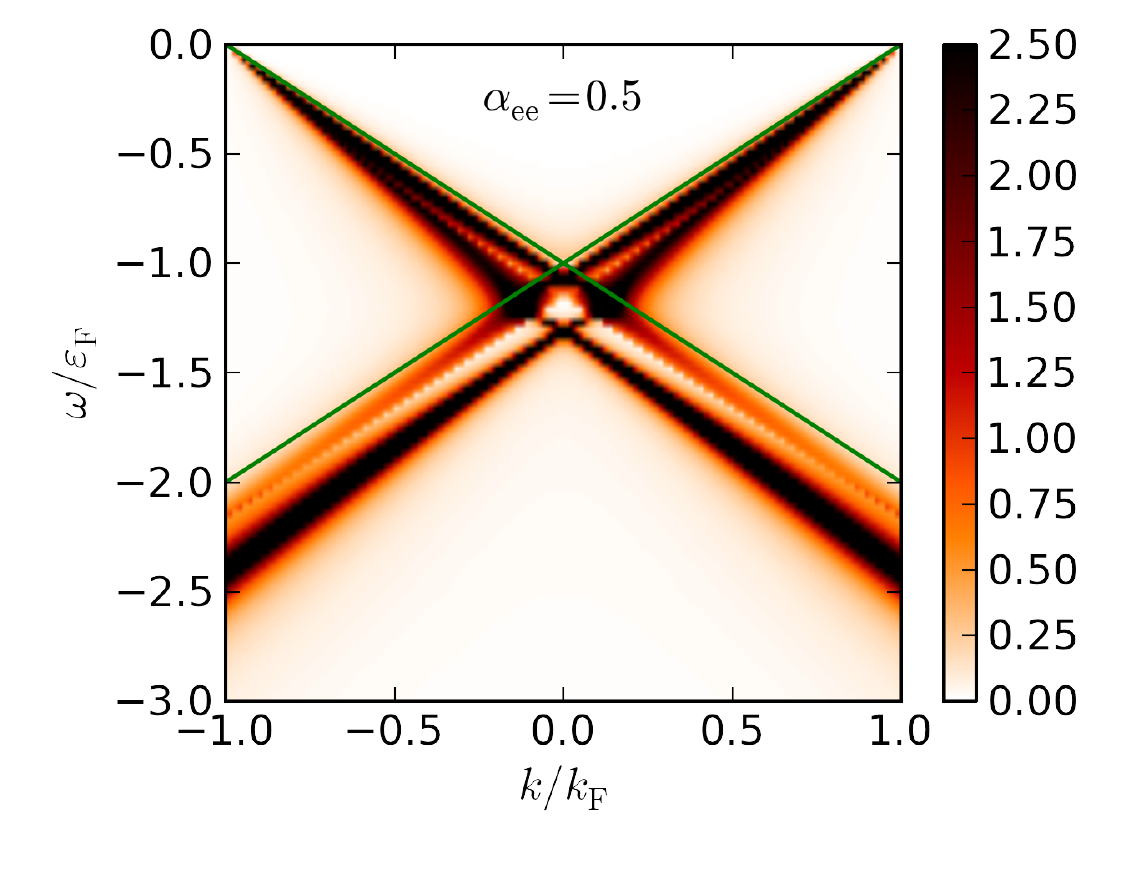}\\
\includegraphics[width=1.00\linewidth]{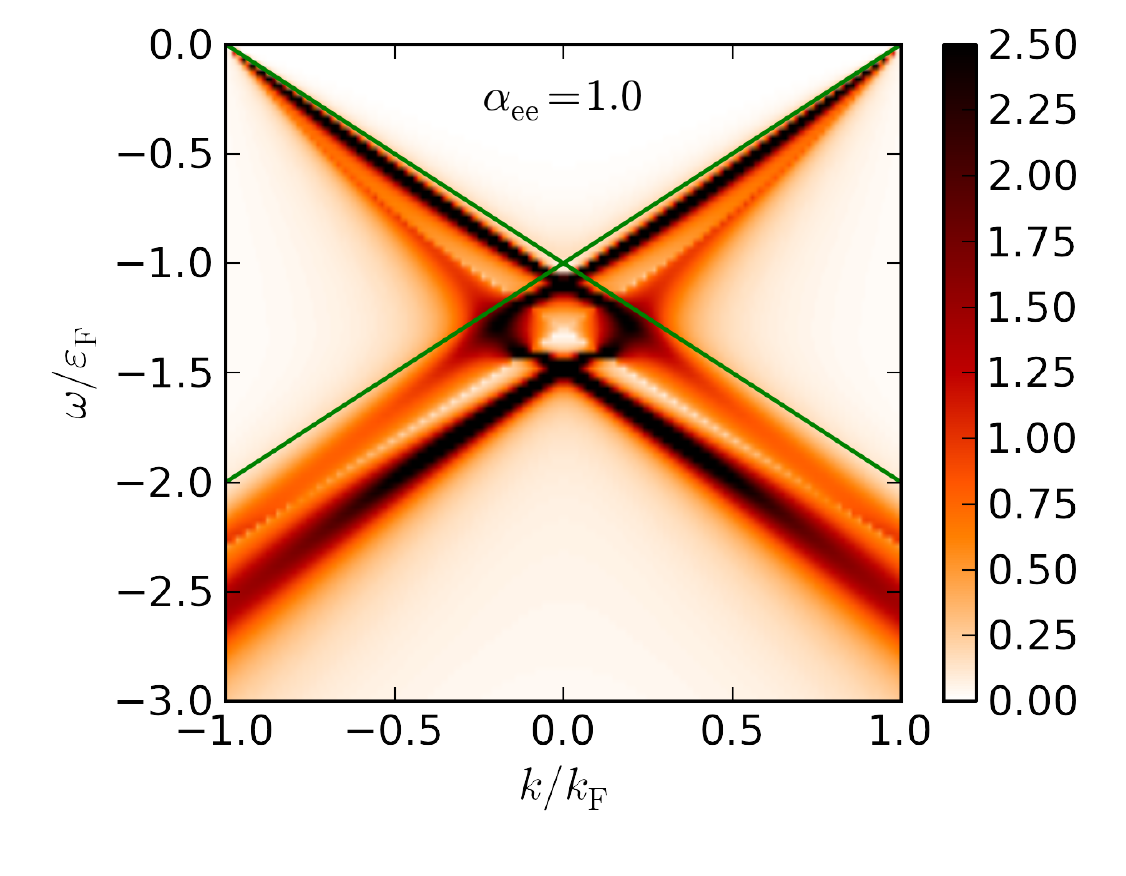}
\caption{(Color online) Same as in Fig.~\ref{fig:sm_1} but for two smaller values of the coupling constant $\alpha_{\rm ee}$.
\label{fig:sm_2}}
\end{figure}

In Fig.~\ref{fig:sm_2} we report plots of the $G_0W$-RPA spectral function ${\cal A}({\bm k}, \omega)$ of the electron gas in G/M for $n = 5.0 \times 10^{13}~{\rm cm}^{-2}$, $d =3~{\rm \AA}$, and two values of the coupling constant $\alpha_{\rm ee}$. Comparing these results with those reported in Fig.~\ref{fig:sm_1}, we clearly see that the soundaron satellites are also robust against changes in the coupling constant. Notice, in particular, that intensity and sharpness of the soundaron peaks increase with decreasing $\alpha_{\rm ee}$. Of course, in the noninteracting $\alpha_{\rm ee} \to 0$ limit these bands smoothly disappear.

\begin{figure}
\centering
\includegraphics[width=1.00\linewidth]{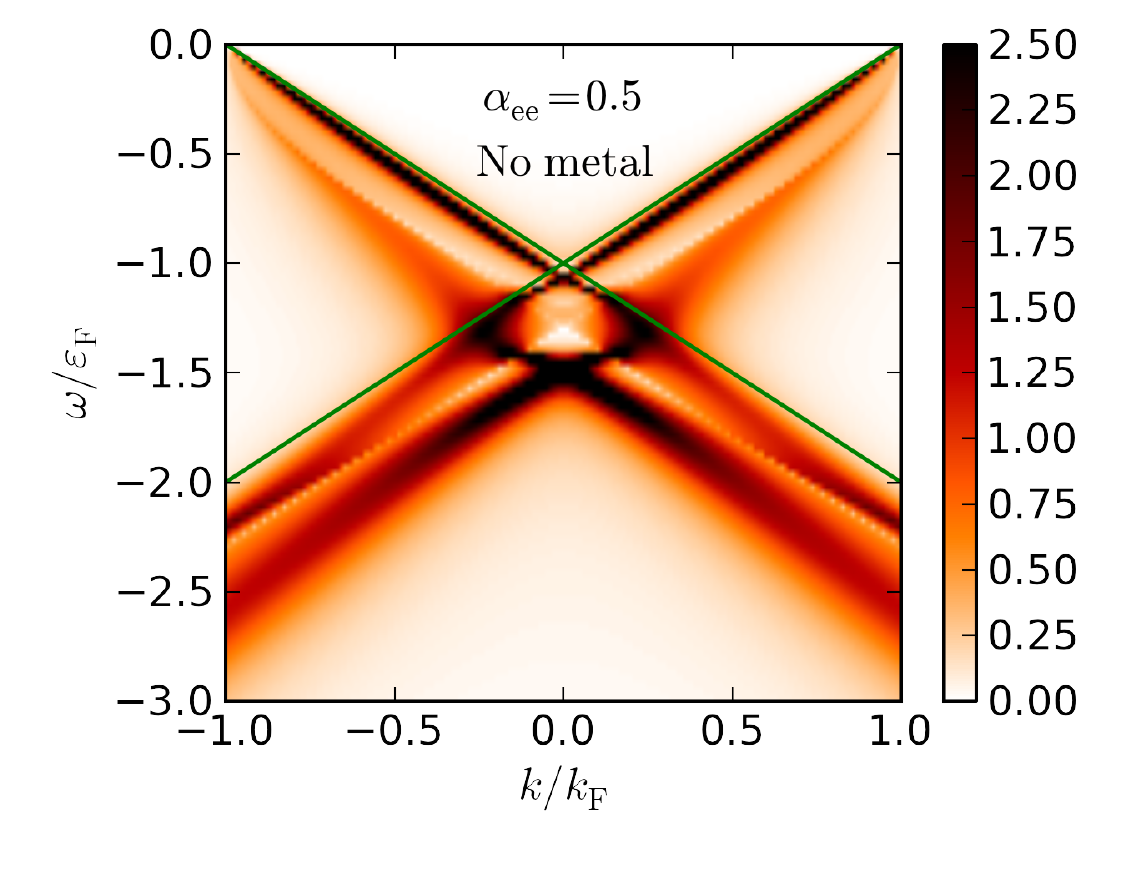}\\
\includegraphics[width=1.00\linewidth]{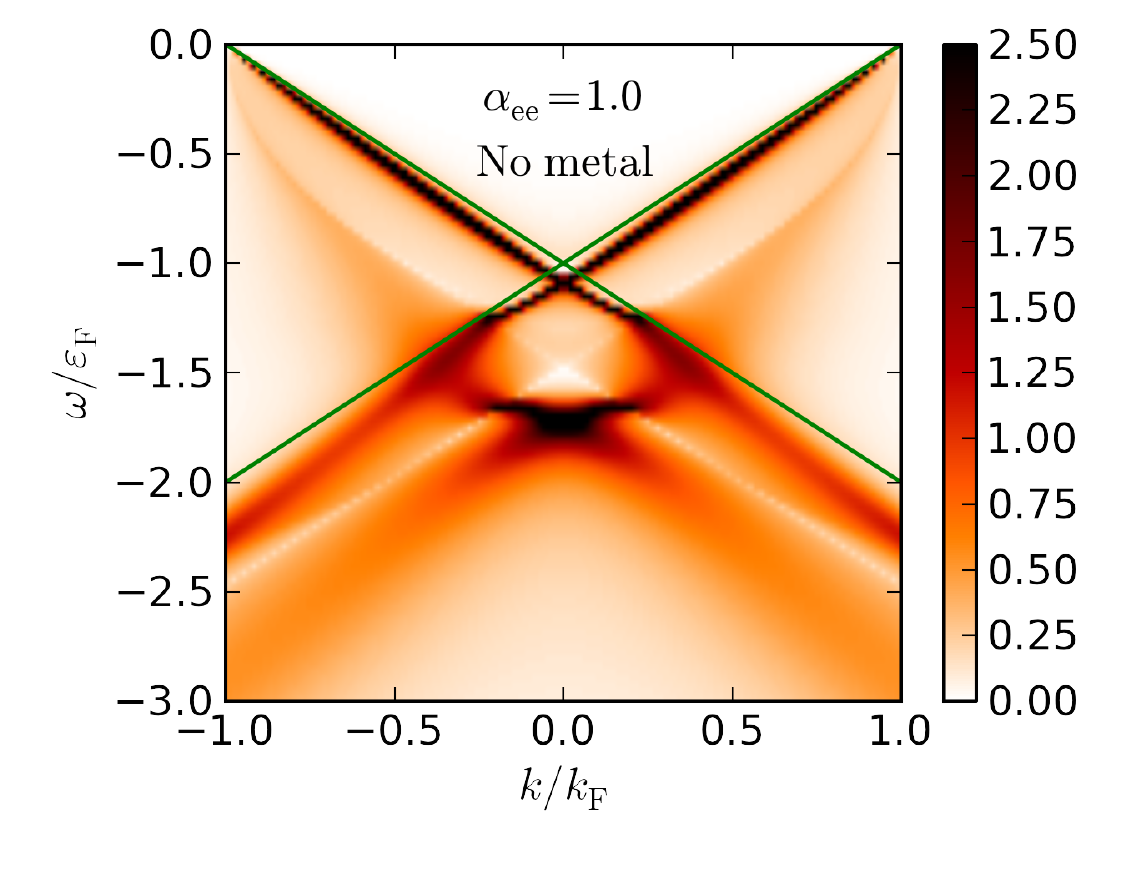}
\caption{(Color online) Same as in Fig.~\ref{fig:four}b) but for two smaller values of the coupling constant $\alpha_{\rm ee}$.\label{fig:sm_3}}
\end{figure}

In Fig.~\ref{fig:sm_3} we illustrate the dependence of the spectral function of an isolated ({\it i.e.} in the absence of a metal gate) graphene sheet on $\alpha_{\rm ee}$. These results should be compared with those reported in Fig.~4b) of the main text. 
Also in this case, the satellite bands due to resonant carrier-plasmon interactions are sharp at moderate-to-intermediate values of $\alpha_{\rm ee}$. This is in agreement with the results reported in the supporting online material of Ref.~\onlinecite{bostwick_science_2010}.

\end{appendix}

\end{document}